\documentclass[referee]{aa} 
\usepackage{graphicx,psfig}
\begin{document}
\title {\bf Magnetars and fossil-field model of magnetic field origin}
\titlerunning{Magnetars and fossil-field model of magnetic field origin}
\author{
 F.K. Kasumov\inst{1}
\and
 A. O. Allahverdiev \inst{1}}

\offprints{F.K. Kasumov}

\institute {Institute of Physics
 of the National Academy of Sciences of Azerbaijan,\\ 33, H.Cavid Str, Baku AZ1143, Azerbaijan Republic\\
\email{(astro@physic.ab.az)}}
\authorrunning{Kasumov et al.}




 \date{Received; accepted}

\abstract
{The evolution and genesis of Anomalous X-ray Pulsars and Soft Gamma ray Repeaters are investigated. The new arguments in favor of magnetar model are found. It is shown that these objects are formed from more massive stars and the fossil-field model is responsible for their high magnetic fields.

}

\maketitle

\section{Introduction}  

In the last years new sources of high energy particles have been discovered. Unlike known classical sources (the Supernovae remnants and radio pulsars) these are new manifestations of neutron stars showing high energy losses and radiating basically in x-ray and gamma ray bands Anomalous X-ray pulsars (AXP) and Soft Gamma-ray Repeaters (SGR).

At present some models explaining their physical nature and evolution (the so-called "fall-back" model responsible for radiation due to accretion, magnetar model the radiation of which is due to magneto dipole mechanism, etc.) are known. Each of these models has its pros and cons (both in comparison of the calculated parameters of neutron stars (NS) with the observational data, and at the comparative analysis of evolutionary behavior of these objects and general notions on the evolution of radio pulsars). All this is exposed lately to intensive research with the purpose of revealing serviceability of one of these models. The given work is devoted to revealing the applicability of fossil-field model to the establishment of the origin of magnetic field of the magnetars. As it has been shown earlier by applying the new test suggested in Allakhverdiev et al., 2005, the magnetars can arise from massive stars and their evolution like that of NS can be described by magneto dipole mechanism (Kasumov \& Allakhverdiev, 2001).

\section {$P-\dot P$ diagram, genesis and evolution of radio pulsars with high magnetic fields and magnetars}  
he main idea of the above-mentioned test consists in comparative study of NS parameters calculated on the basis of available models and tracks of evolution with the observational data on the  $P-\dot P$ diagram, constructed on the basis of most precisely known parameters of pulsars: period P and its derivative $\dot P$. With this purpose the comparison of the age of pulsars calculated from different models with the space-kinematical age of these objects, which are the closest to real values of this parameter, has been carried out. Under the kinematic age of pulsars the value obtained from average velocities of these objects and their distance from the Galactic plane ($|Z|$)) is meant. Since the number of pulsars with known velocity measured by using the proper motion of these objects is not more than 10-15 $\%$ of their total number, average values of velocities of these objects following from their evolutionary-genetic reasons are usually used, that justifying itself very well at statistical researches (Guseinov et al., 2003). As the velocities of pulsars obtained during the explosion of the Supernova (both due to asymmetry of explosion, and in the result of a pair disruption in the binary system) do not exceed the escape velocity of these objects from the Galaxy, and their age $\sim 10^{7}$ years, then there should be a linear dependence between the real age of pulsars and their distance from the Galactic plane.

In order to see more precisely the age difference, the pulsars used in the analysis have been divided into two groups on $|Z|$-coordinate: the young ones with $|Z|< 100 $ pc and the oldest ones $|Z|> 300 $ pc. The data on distances are taken by us from (Guseinov et el., 2003) and the value of average velocities - from (Manchester et al., 2005).

\begin{figure}
\centering
\includegraphics[width=13 cm]{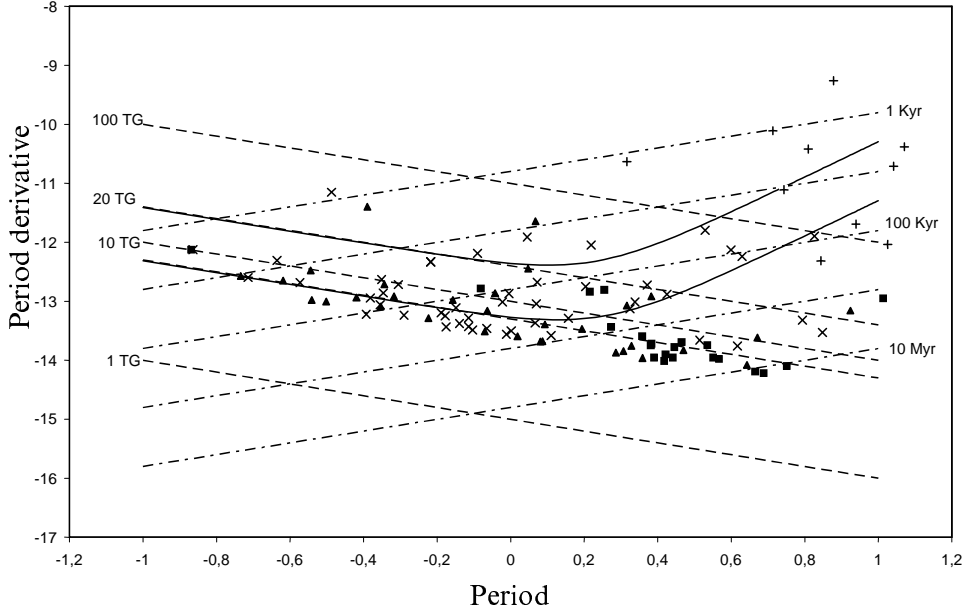}
\caption{ $P-\dot P$ diagram for pulsars with $B>5\times 10^{12}\, G$ and magnetars.  Dotted lines  correspond to the lines of constant characteristic age $\tau$, strokes - dashed lines - constant magnetic field $B$. Solid-curved lines mark evolutionary track described by combine model (6). Other symbols see in the text.  }
\label{fig1}
\end{figure}

As it is seen from the $P-\dot P$ diagram (see Fig. 1) constructed only for pulsars with high magnetic fields ($B \ge 5 \times 10^{12}$ Gs) and magnetars, the use of these objects is stipulated by the above purpose: to study their evolutionary behavior, and also to establish the origin of the magnetic fields of magnetars. It is suggested that the origin of magnetic fields of both these objects is the same. The dashed lines on the diagram show the lines of a constant magnetic field $B$(i.e. tracks of evolution of pulsars with constant magnetic field following from expression   $ B= 3.2 \times 10^{19} (P\dot P)^{1/2}$(Alpar et al.,2001) at the values of a magnetic field $B = 100, 10, 1 $ TGs, 1 TGs (1 TGs $= 10^{12} $ Gs). Strokes - dashed lines correspond to constant characteristic age $\tau =0.5 (P/\dot P)^{1/2} $, with values at $\tau =   10^{3} , 10^{4}, 10^{5}, 10^{6}, 10^{7} $ years, accordingly. Two curves on the diagram describe the evolution tracks of pulsars according to combined (magneto dipole + accretion) models. The combined model is meant to be the model in which joint action of the magneto dipole and accretion or propeller mechanisms (Alpar et al.) are taken into account.

According to this model the growth of the period of pulsar occurs under the influence of both mechanisms according to expression

$$ I\Omega\dot\Omega=-\beta {\Omega}^{4} -\gamma  \eqno (1)  $$
where $I$ is an inertia momentum of the pulsar rotating with angular velocity $\Omega$. Here the first member describes magneto dipole, and the second one propeller mechanism of loss. The NS will continue to act as a radio pulsar as long as the fall-back disk does not protrude into the light cylinder. The torque due to this disk can be estimated as:
	$$	N_{disk}= -\gamma/\Omega  \eqno (2) $$ 
 Here $$ \gamma= 2\times 10^{31} \dot {M}_{10}  \qquad \mbox  {\it erg/sec} \eqno (3) $$
 Where $ \dot{M} $
   is the accretion rate in units of $ 10^{10}$ g/sec.
The rate of energy loss due to magnetodipole radiation is given by
	$$ \dot {E}_{dipol}=- {B_{\bot}}^{2} R^{6}\Omega^{4} /6\times c^{3} \eqno (4) $$ 

Value  $ \beta $ in this expression is determined as
$$ \beta=6.17 \times 10^{27}(B_{\bot,12})^{2}R_{6}^{6} \eqno  (5)$$ 
where $ B_{\bot,12}$ is the component of a magnetic field $Â $ (in units of $ 10^{12}$ Gs) perpendicular to the rotation axis, $ R_{6}$ is the NS radius in units of $10^{6}$ cm. Hence, the change of the pulsar period can be determined from (1) according to expression
$$ \dot{P} = \left(4  \pi^{2}\beta/I \right)P^{-1} + \left(\gamma/4 \pi^{2}I \right)P^{3}  \eqno (6)$$

When constructing the evolution tracks according to the combined model the following parameters were used: the upper curve $B =20 $ TGs, $ \dot{M} =10^{11}  $ g/sec, the lower curve - $B =7 $ TGs, $ \dot{M} =10^{10}  $ g/sec.

Designations of objects are as follows: magnetars - $ \bf {+} $, radio pulsars: with $ |Z|< 50 $ pc - $\bf {\triangle} $, (where $|Z|$ is the distance of a pulsar from the Galactic plane), $50 $ pc $< |Z| < 100 $pc  -  x,  $|Z| > $ 300   pc  - ``black square''.

In order to clearly see the tendency mentioned in the introduction, the first group of pulsars ($|Z|< 100 $ pc) has been subdivided into two subgroups with $|Z|< 50 $ pc and  $50$ pc $<|Z| <100$ pc, i.e. the difference of progenitor star masses have been considered.

As it is clearly seen from Fig. 1 the number density ratio of pulsars with $ |Z|< 50 $ pc to density of pulsars with $50 $pc $<|Z|<100$ pc with the growth of magnetic field continuously increases from 1,4 and 1,7 ( ${ B =  10^{12}}$ Gs) up to 2,6 (at ${B  =  10^{13} }$ Gs, on the tracks of both evolution models) i.e. the tendency of growth of a magnetic field of objects located near the Galactic plane is observed. In other words, magnetic fields of the pulsars formed from massive stars ($|Z| <50$ pc) are higher than the magnetic fields of the pulsars which have arisen from less massive stars. Thus, the nearer pulsars to the Galaxy plane, the more their magnetic fields. It confirms the general idea about the formation of radio pulsars with high magnetic fields and perhaps magnetars from more massive stars ($M   \ge 40 Ì_{\odot} $, here  $ M_{\odot}$ is the Sun mass) (Kasumov \& Allakhverdiev, 2001).

At the realization of the magneto dipole mechanism for the explanation of the origin of high magnetic fields two scenarios are discussed. According to the first of them a part of OB stars with rather high magnetic field ($ \sim 3\times 10^{3}$ Gs) during the evolution and transformations into a compact NS at the conservation of a magnetic flux, their magnetic field can increase as a result of change of initial and final radii of object (as $ R^{-2})$). It is interesting to note, that magnetic fluxes of two massive stars ($M\ge  40 Ì_{\odot}$) for which the magnetic fields are measured, can be compared to the magnetic flux of the magnetar SGR0806-20 equal to $ 5\times 10^{27} $ Gs/$sm^2$  (Ferrario \&Wickramasinge, 2006). Moreover, recent observations have revealed two massive stars Ae/Be that have just entered to the main sequence (MS), (these are W601 and OI 201 with $ M \ge 10 M_{\odot}$) on Hertzsprung-Russell diagram which by the spectral polarimetric observation data have the estimated magnetic fields -$ \sim 3\times 10^{3} $ Gs (Alecian et al.,).

According to the second scenario during the collapse of massive stars and the transformation of their nucleus into a proto-neutron star as a result of fast rotation (the period $\le 3$ msec) and the action of the dynamo-mechanism during the short time ($\le 10 - 100$ sec) the magnetic field of nucleus can sharply increase up to $10^{13} - 10^{14} $  Gs.

In order to establish which of these two scenarios play a key role in the emergence of such high magnetic fields of these objects, we shall again refer to Fig. 1. As it is seen from this figure, the number density of objects does not grow with the increase of the period of pulsars both for the objects with $|Z| <50 $pc, and with $50 $pc$ < |Z| <100$ pc, and the pulsars are almost evenly distributed on the period intervals that contradicting the dynamo-mechanism model. In fact, according to this model the initial periods of pulsars should be $\le 3 $  msec and according to theoretical calculations that take into account various types of losses of the NS rotation moment, during a short period of time ($\sim 10 - 100 $ sec) these periods should grow up to the values close to those of magnetars ($ \sim 5 $ seconds). In that case a tendency of growth of the number of objects should be observed in the field of the large periods.

However, as it is seen from Fig. 1., the number of pulsars within the periods 0.1 sec < P < 0.63 sec, 0.63 sec < P < 1.58 sec and 1.58 sec < P < 6.3 sec, is accordingly 24, 30, 20, i.e. the distribution of pulsars is almost regular and the expected tendency in case of the action of the dynamo-mechanism is not observed. Certainly, such tendency should be only if the theoretical calculations, especially those connected to the rotation moment loss, would proceed for the period $\sim 10 - 100 $ sec as it is described in Vink \& Kuper. On the other hand, as it follows from Fig. 1 the ratio of numbers of objects with the higher magnetic fields to the number of objects with smaller ones ($ n_{B>B_{o}}/n_{B\le B_{o}}$ ) with the increase of the period shows the increase, expected in this case, for several times (of course, at rather weak statistics, approximately 20 objects). This circumstance does not allow excluding completely the possibility of realization of the dynamo-mechanism.

If the time of losses, or correspondingly the growth of the period of a pulsar stretches for a longer period of time, for example, due to process of permanently acting glitches, as it was offered by Lin \& Zhang (2004), such tendency could not be observed. However, as it has been shown by us earlier (Kasumov et al., 2006), this mechanism of formation of magnetars from the radio pulsars with high magnetic fields and initially small periods of rotation, also appeared inefficient. But all this certainly does not exclude an opportunity of creating other mechanisms of the long process of the period growth (but not more than $10^{4}-10^{5}$ years, i.e. the age of magnetars), as, for example, unique cases when separate pulsars get into dense interstellar clouds in which there can be rather effective rotation moment losses. However, we consider it to be artificial, since in that case either all magnetars would have appeared to be in a dense environment and would have Wind nebulas or they would be situated in the remnants of supernovae (SNR) with high density of matter. Actually, the observational data do not show such picture (only a few magnetars are connected to SNRs which do not differ from usual SNRs neither by their power, nor by the density of matter (Vink \&Kuper, 2006).

\section {Discussion.}

As it has been shown earlier (Kasumov \& Allakhverdiev, 2007), as well as according to the above analysis, the origin of high magnetic fields of radio pulsars and magnetars from some magnetized massive OB stars is confirmed. However, at the same time the suitability of these to possible mechanisms (fossil-field or the dynamo-mechanism) still remains a subject of discussion. The analysis carried out above and the results of this work show that the realization of the fossil-field scenario is more acceptable and agrees more with observations than that of a dynamo mechanism.

In fact, if the dynamo mechanism acted effectively during the formation of NS and provided the newly born magnetars with fields $10^{15} $Gs then the SNRs connected with magnetars should possess the high residual magnetic fields and be more energetic. However, in reality, the magnetic field and energy losses at the birth of these objects do not differ from other SNRs and correspond to the standard estimations of the latter [8]. In its turn, it makes more attractive the fossil-field model in an explanation of the forming of high magnetic fields in magnetars.

Actually, the massive star with $ M \sim  40 Ì_{\odot}$ with an initial magnetic field on MS $ \sim 3\times 10^{3}$ Gs, being compressed to the sizes of NS can explain observable magnetic fields of magnetars ($ \sim 10^{15} $ Gs) only due to the conservation of magnetic flux. Moreover, in this case a massive star should have such speed of rotation, in order to provide, as a result, the observable periods of magnetars ($\sim 5-12$  sec), both in the case of conservation of the rotation moment, and taking into account the possible losses of the rotation moment during the evolution.

The magnetars can, certainly, arise from the proto-neutron nucleus of massive stars by the dynamo-mechanism as well. However, for their transformation into the magnetars, i.e. for the increase of the initial periods of proto-neutron nucleus ($\sim  3 $ msec) up to values characteristic for magnetars (5-12 sec), the certain braking mechanisms providing  $ \sim 70 \%$  of loss of the rotation moment (for example, magneto dipole, gravitational, etc.) are necessary.

As it has been shown by various authors (Vink \& Kuper, 2006, Lin \& Zhang, 2004, Allakhverdiev et al., 2005, Heyl, 2005), taking into account the possible mechanisms of loss leads to difficulties in the realization of this scenario. One of such mechanisms could be that proposed in (Lin \& Zhang, 2004) for newly born young pulsars with the periods   $\sim $ msec, namely the mechanism of prolonged and repeating glitches with relatively high values. But as it was already stated above, the action of this mechanism was not confirmed by the observational distribution and evolution of such pulsars in the  $P-\dot P$ diagram (Kasumov et al., 2006).

On the other hand, the large losses of the rotation moment can lead to losses of the magnetic moment too, as a result this will lead to delay and restriction of the growth of magnetic field of the formed NS which may not reach the values characteristic for magnetars.

On the contrary, in fossil-field models such contradictions are not present and as it was mentioned above, the magnetic flux of massive stars is in good agreement with an observable fluxes for magnetars.

Finally, in the recent review (Mereghetti, 2008) devoted to magnetars, a number of arguments against availability of the dynamo-mechanism is in detail discussed. One of them is connected to the necessity of high spatial velocities ( $ \sim 1000 $  km/s) for magnetars at the action of this mechanism. However, the presence of three magnetars reliably connected with the SNRs (magnetars situated in their center), and the proper motion measurements of the magnetar XTE J1810-197 provide the velocity  $ \le 180 (d/3êïê)$ $\le 180(d/{3\,kpc}) $  km/sec, i.e. the value characteristic for usual pulsars that testify to the opposite. On the other hand, the location of magnetars near the young star associations and the absence of considerable differences between the SNRs connected to them from the others by the energy of explosion also should be considered as arguments against the action of a dynamo mechanism.

We should also note that the similar statement of the problem with a slightly different method of the analysis has been carried out independently from us quite recently in (Ferrario \&Wickramasinge, 2008) the conclusions of which have completely coincided with the results obtained in the given work.

\clearpage

\end{document}